# Quantitative modeling of degree-degree correlation in complex networks

A. Niño[1] and C. Muñoz-Caro[2]

[1] Center for Complex Networks Research, Northeastern University, Boston, Massachusetts 02115, USA, on leave from[2]

[2] E. S. Informática de Ciudad Real, Universidad de Castilla-La Mancha, 13004 Ciudad Real, Spain

**Abstract**

This paper presents an approach to the modeling of degree-degree correlation in complex networks. Thus, a simple function, $\Delta(k', k)$, describing specific degree-to-degree correlations is considered. The function is well suited to graphically depict assortative and disassortative variations within networks. To quantify degree correlation variations, the joint probability distribution between nodes with arbitrary degrees, $P(k', k)$, is used. Introduction of the end-degree probability function as a basic variable allows using group theory to derive mathematical models for $P(k', k)$. In this form, an expression, representing a family of seven models, is constructed with the needed normalization conditions. Applied to $\Delta(k', k)$, this expression predicts a nonuniform distribution of degree correlation in networks, organized in two assortative and two disassortative zones. This structure is actually observed in a set of four modeled, technological, social, and biological networks. A regression study performed on 15 actual networks shows that the model describes quantitatively degree-degree correlation variations.

**PACS numbers:** 02.50.-r, 89.75.Da, 89.75.Fb, 89.75.Hc

**I. INTRODUCTION**

Significant effort has been made in the recent years in the study of the structure and behavior of complex social, biological, and technological networked systems [1]. In any network, the degree probability distribution [1, 2] is used as the most basic topological characterization. However, as pointed out by Newman a decade ago [3] networks can -and real networks usually do- exhibit a certain pattern, a correlation, among the degrees of nodes connected together. This degree-degree correlation is not accounted for by the degree distribution. Newman [3] introduced the terms assortative and disassortative to denote networks where nodes of similar or different degree tend to connect together. Newman also provided the first clear example of the importance of degree correlation in network behavior by showing that resilience (the ability to maintain network connectivity) is enhanced in assortative networks [3].

Resilience is an expression of robustness, understood as the ability of a network to keep its functioning under failure of some of its components. This topic has been addressed on specific systems such as the Internet [4-6]. In addition, several works have considered how biased (degree-dependent) percolation provides strategies to turn fragile networks into robust ones [7, 8]. In this context, different studies on robustness of networks have considered the key role played by degree correlation. Thus, Serrano et al. [9] have found that structural constrains prevent the existence of uncorrelated weighted networks. On the other hand, it has been shown [10] how resilience to damage is a function of the dependence of the percolation threshold on degree correlation. In addition, Goltsev et al. [11] have used two models of correlated networks, with strong assortative and disassortative mixing, to show that correlation changes not only the percolation threshold but also the critical behavior at the percolation point. Recently, Tanizawa et al. [12] have analytically evidenced the role played by degree correlation in enhancing the robustness of the "onionlike" networks introduced by Schneider et al., [13], where nodes of very close degree are connected together (the layers of the "onion").

Percolation and robustness are not the only subjects affected by correlation. Spreading phenomena, a term that applies to the propagation in networks of different entities such as epidemics, rumors, opinion formation, or computer viruses, are also influenced by degree correlation. In this context, epidemic spreading has been extensively studied [14-18] and can be considered the canonical example for the whole set of spreading phenomena. These studies show that epidemic dynamics is dependent on degree correlation. Also, it has been found that in general networks the epidemic threshold is given by the inverse of the largest eigenvalue of the connectivity matrix, which is defined by the conditional probability of a node of degree k' to be connected to a node of degree k [15, 16]. However, no epidemic threshold exists in scale-free networks with node to node correlation [14].

Several other examples of the relevance of degree correlations can be found. Thus, it has been shown that transport processes in scale-free networks exhibit a great dependence on degree correlation [19]. Recently, the influence of degree correlation in synchronization within scale-free networks has been considered. The results show that optimal synchronization needs a specific degree of assortativity [20].

Degree correlation also affects the controllability of dynamical networks. In particular, simulation studies have shown that disassortative mixing enhances controllability [21]. Degree correlation is also a key factor in dynamic models of network growing, as those able to build a network with a predefined behavior [22].

Different ways of describing degree correlation in complex networks have been used. Thus, the conditional probability of a node of degree k' to be joined to a node of degree k, *P(k' | k)*, was introduced in [23]. In that same work the average nearest neighbors degree (ANND) function was presented. Callaway et al. [24] considered the concept of joint degree probability distribution of nodes of degree k' and k, *P(k', k)*, as a correlation measure. Later, Newman introduced the Pearson correlation coefficient of the degrees of nodes of the edges, $r_c$, as a global index to quantify the assortative or disassortative character of a network [3]. Excellent reviews of the topic are available [16, 25].

Several ad hoc models can be found in the literature for describing degree correlation between different network models. Thus, an expression for *P(k' | k)* is used in refs. [10, 26] to describe the passage from uncorrelated to correlated networks as a function of a single parameter. Similarly, Tanizawa et al. [12] have used an expression for P(k', k) that allows changing from a random regular to a completely random network as a function of a control parameter.

However, a question remains. Given a real network, how can we describe accurately the existing degree correlation? In turn, what is the simplest expression we can use to describe accurately degree correlation in networks? These are necessary questions if we want to properly understand, and in the last instance control, the behavior of actual systems.

In this work, we present an approach to the development of degree correlation functions for general complex networks. The functions are useful for building complex networks with a predefined assortative or disassortative behavior or for describing quantitatively and analytically the correlation observed in given networks. The proposed approach is contrasted against different real systems.

## II. DEGREE CORRELATION MODEL

To analyze degree correlation in networks, Newman [3, 25] introduced the concept of assortativity and disassortativity through the $r_c$ degree correlation coefficient defined as,

$$r_c = \frac{\sum_{k'}\sum_{k} k'k[P(k',k) - q(k')q(k)]}{\sum_{k} k^2 q(k) - [\sum_{k} k q(k)]^2} \tag{1}$$

Here, $q(k)$ represents the end-degree probability distribution corresponding to the probability of having an end node of degree $k$ in an edge. This is the $q_k$ distribution considered in ref. [27]. We will use here the $q(k)$ notation.

The denominator in equation (1) is just a normalization factor. It is the numerator which defines the character of the network. Considering that $q(k')·q(k)$ is the joint probability distribution for the uncorrelated case, we have assortativity when $r_c>0$ (joint probability higher than in the uncorrelated case), neutrality for $r_c=0$, or disassortativity when $r_c<0$ (joint probability smaller than in the uncorrelated case). However, $r_c$, and even the ANND function [23], only provide partial pictures of the degree correlation distribution.

Different ways to measure correlations deviations can be devised. Thus, Maslov and Sneppen [28] considered the deviation of *P(k', k)* from a randomized version of the same network. In this form, they define a correlation index, *Z(k', k)*, as the quotient of this deviation and the standard deviation of multiple realizations of the random network. To quantify degree correlation variations within a network, we adopt here a direct approach, considering a function, Δ(*k', k*), measuring the deviation from the uncorrelated case for given *k'* and *k* values:

$$\Delta(k', k) = P(k',k) - q(k') \cdot q(k) \tag{2}$$

Equation (2) shows that $\Delta(k', k) = \Delta(k, k')$ and $-1 \leq \Delta(k', k) \leq +1$. $\Delta(k', k)$ directly expresses the inner hierarchical distribution of degree correlations in a network. A graphical representation of $\Delta(k', k)$ could be used to visualize the network correlation behavior as a function of $q(k')$ and $q(k)$ or $k'$ and $k$. By analogy with Newman's $r_c$ definition [3, 25], $\Delta(k', k) > 0$ represents an assortative relationship between degrees $k'$ and $k$, whereas $\Delta(k', k) < 0$ corresponds to a disassortative one. For the uncorrelated case we have $\Delta(k', k) = 0$.

However, the question is how to develop a functional form describing degree correlation variations in an arbitrary network. The starting point is to consider two events, *A* and *B*, defined in the same probability space, with probabilities *P(A)* and *P(B)*, respectively. Thus, we have [29]

$$P(A|B) = \frac{P(A,B)}{P(B)} \qquad (3)$$

where *P(A|B)* is the conditional probability of *A* given *B*, and *P(A, B)* is the joint probability of *A* and *B*. These distributions satisfy the relationships:

$$\begin{aligned} P(A|B) &\neq P(B|A), \\ P(A,B) &= P(B,A) \end{aligned} \qquad (4)$$

The symmetry exhibited by the joint distribution makes it more suitable for describing degree correlation. Thus, we have for independent events,

$$P(A,B) = P(A) \cdot P(B) \qquad (5)$$

Now, in the spirit of the virial expansion used in the study of real gases [30], let us define an $\alpha$ coefficient as,

$$\frac{P(A,B)}{P(A) \cdot P(B)} = \alpha \qquad (6)$$

Therefore, $\alpha=1$ for independent events and $\alpha \neq 1$ otherwise. Since for independent events *P(A, B)* depends on *P(A)* and *P(B),* we can consider that, in the general case*: P(A, B)=f [P(A), P(B)]*. In fact, models of growing networks show that the joint degree probability distribution exhibit a dependence on the degree distribution [31, 32]. This makes $\alpha$ depend on *P(A)* and *P(B)*. Thus, defining *X=P(A)*, *Y=P(B),* we have $\alpha=\alpha$*(X, Y)*. This approach has three advantages over the equivalent case of $\alpha$ expressed as a function of A and B. First, we can work with P(A) and P(B) independently of its specific form. Second, we only need to handle P(A) and P(B) without explicit consideration of A and B. Third, the use of P(A) and P(B) brackets the range of X and Y to the [0, 1] interval.

On the other hand, the interchange symmetry of the joint distribution, equation (4), suggests defining two operators acting on an arbitrary *f(X, Y)* function. The first one is the identity operator, $\hat{E}$, and the second one is an interchange operator, $\hat{I}$. The effect of these operators is illustrated in equation (7).

$$\begin{aligned} \hat{E} f(X,Y) &= f(X,Y) \\ \hat{I} f(X,Y) &= f(Y,X) \end{aligned} \qquad (7)$$

These two operators define an interchange group, $G_2$, isomorphic to the $C_s$ point group [33]. Its character table is shown in Table I.

Considering the interchange symmetry of *P(A, B)* and *P(A)·P(B)*, equation (6) shows that $\alpha$*(X, Y)* must also be invariant under the $\hat{I}$ operation. Let us now expand $\alpha$*(X, Y)* as a Taylor series in *X* and *Y* around an arbitrary point $X_0$, $Y_0$. To fulfill the symmetry conditions, we apply the projection operator of the $a_1$ irreducible representation, see Table I, to the Taylor expansion. Thus, we obtain a symmetry-adapted Taylor expansion as,

$$\alpha(X,Y) = \hat{P}_{a1}\left[\sum_{l=0}^{\infty}\sum_{m=0}^{\infty}\frac{1}{(l+m)!}\left(\frac{\partial^{l+m}\alpha(X,Y)}{\partial^l X\,\partial^m Y}\right)_0 \Delta X^l \Delta Y^m\right] \quad (8)$$

Applying the projection operator and expanding the powers of $\Delta X$ and $\Delta Y$ we have,

$$\alpha(X,Y) = \sum_{l=0}^{\infty}\sum_{m\geq l}^{\infty} a_{lm}(X^l Y^m + X^m Y^l) \quad (9)$$

To apply equation (9) to a network, we must realize that the probability distribution corresponding to *P(A)* and *P(B)* is not the degree probability distribution, *P(k)*, but the *q(k)* distribution considered in ref. [27]. *q(k)* and *P(k)* are related by [16, 25, 27],

$$q(k) = \frac{kP(k)}{\sum_k kP(k)} = \frac{kP(k)}{\langle k \rangle} \quad (10)$$

In fact, it is easy to prove (see equations (7) and (8) in ref. [16]) that only *q(k)* satisfies the product rule given in equation (3).

With the previous considerations, and substituting equation (9) in equation (6), we get

$$P(k',k) = \sum_{l=0}^{\infty}\sum_{m\geq l}^{\infty} a_{lm}\left[q(k')^{l+1} \cdot q(k)^{m+1} + q(k')^{m+1} \cdot q(k)^{l+1}\right] \quad (11)$$

Equation (11) allows expression of the joint probability distribution as a function of probability distributions, not of node degrees. Having actual data for *P(k', k)*, *q(k')*, and *q(k)* we can use equation (11) to obtain a functional expression for *P(k', k)* by applying a multilinear regression on the $a_{lm}$ coefficients. In any case, the resulting expressions must be properly normalized by fulfilling the conditions [16, 25],

$$\sum_{k'=1}^{k_m} \sum_{k=1}^{k_m} P(k',k) = 1,$$
$$\sum_{k'=1}^{k_m} P(k'|k) = 1 \qquad (12)$$

Equation (12) corresponds to a network, where every node is connected at least to another one and where the maximum degree, $k_m$, is finite. Therefore, the conditions in equation (12) are a consequence of the following property [16, 25],

$$\sum_{k'=1}^{k_m} P(k',k) = q(k) \qquad (13)$$

Equation (13) tells us that the coefficients in equation (11) are not independent. They must satisfy the condition,

$$\sum_{l=0}^{\infty} \sum_{m \geq l}^{\infty} a_{lm} [I_{l+1} \cdot q(k)^{m+1} + I_{m+1} \cdot q(k)^{l+1}] = q(k) \qquad (14)$$

where the $I_n$ notation stands for the indices

$$I_n = \sum_{k=1}^{k_m} q(k)^n \qquad (15)$$

Obviously, $I_1=1$, and $1 > I_n > I_{n+1}$ in any other case. Equation (14) implies automatic fulfillment of the normalization conditions in equation (12).

The simplest approach is obtained by limiting the expansion in equation (11) to the terms with l+m≤1. This yields a two terms expression. However, by using equation (14), it is easy to show that no expression of the form *P(k', k) =a·q(k')·q(k)+b·f(k',k)*, where the f function does not contain linear terms in q(k), satisfies equation (13).

We can build a flexible expression by limiting the series expansion to third order terms [*i. e.,* l, m ≤ 2 in equation (11)]. Thus, we obtain

$$P(k',k) = a \cdot q(k') \cdot q(k) + b \cdot [q(k') \cdot q(k)^2 + q(k')^2 \cdot q(k)] + c \cdot q(k')^2 \cdot q(k)^2 + d \cdot [q(k') \cdot q(k)^3 + q(k')^3 \cdot q(k)] + e \cdot [q(k')^2 \cdot q(k)^3 + q(k')^3 \cdot q(k)^2] + f \cdot q(k')^3 q(k)^3$$

(16)

Equation (13), allows fixing three parameters. Here, we select to keep the three parameters corresponding to the most significant terms, a, b and c. So, we obtain,

$$d = \frac{[1-a-bI_2]}{I_3}$$
$$e = -\frac{(b+cI_2)}{I_3}$$
$$f = \frac{[2bI_2+cI_2^2+a-1]}{I_3^2}$$

(17)

Therefore,

$$\begin{aligned}P(k',k) =& \\ & a \cdot [q(k') \cdot q(k) - (1/I_3) \cdot (q(k') \cdot q(k)^3 + q(k')^3 \cdot q(k)) + (1/I_3^2) \cdot q(k')^3 q(k)^3] \\ &+ b \cdot [(q(k') \cdot q(k)^2 + q(k')^2 \cdot q(k)) - (1/I_3) \cdot (q(k')^2 \cdot q(k)^3 + q(k')^3 \cdot q(k)^2) \\ &\quad - (I_2/I_3) \cdot (q(k') \cdot q(k)^3 + q(k')^3 \cdot q(k)) + (2I_2/I_3^2) \cdot q(k')^3 q(k)^3] \\ &+ c \cdot [q(k')^2 q(k)^2 - (I_2/I_3) \cdot (q(k')^2 \cdot q(k)^3 + q(k')^3 \cdot q(k)^2) + (I_2^2/I_3^2) \cdot q(k')^3 q(k)^3] \\ &+ (1/I_3) \cdot (q(k') \cdot q(k)^3 + q(k')^3 \cdot q(k)) - (1/I_3^2) \cdot q(k')^3 q(k)^3\end{aligned}$$

(18)

Equation (18) satisfies equation (13) for any combination of the independent term with the a, b and c terms. Thus, equation (18) actually represents a family of seven models corresponding to the possible combinations of the a, b and c contributions. In these models, the effect of the network topology is introduced through the $I_2$ and $I_3$ indices. Clearly, equation (18) satisfies d=e=f=0 for a=1, b=0 and c=0.

The degree-degree correlation within a network can be described using the *Δ(k', k)* function. Thus, using equation (18),

$$\Delta(k', k) =$$
$$(a - 1) \cdot [q(k') \cdot q(k) - (1/I_3) \cdot (q(k') \cdot q(k)^3 + q(k')^3 \cdot q(k)) + (1/I_3^2) \cdot q(k')^3 q(k)^3]$$
$$+ b \cdot [(q(k') \cdot q(k)^2 + q(k')^2 \cdot q(k)) - (1/I_3) \cdot (q(k')^2 \cdot q(k)^3 + q(k')^3 \cdot q(k)^2)$$
$$- (I_2/I_3) \cdot (q(k') \cdot q(k)^3 + q(k')^3 \cdot q(k)) + (2I_2/I_3^2) \cdot q(k')^3 q(k)^3]$$
$$+ c \cdot [q(k')^2 q(k)^2 - (I_2/I_3) \cdot (q(k')^2 \cdot q(k)^3 + q(k')^3 \cdot q(k)^2) + (I_2^2/I_3^2) \cdot q(k')^3 q(k)^3]$$

(19)

To analyze this expression, we represent the (a-1), b and c terms as cases a), b) and c) of Figure 1, using generic values $I_2$=0.2 and $I_3$=0.02. We observe that the three terms exhibits a similar pattern. In particular, the (a-1) and c terms show the same behavior, with four separated zones clearly apparent in the graphs. Using equation (19) it is straightforward to show that the (a-1) and c terms cancel for *q(k)* or *q(k')* = α, with α=√$I_3$, in the first case, and α= $I_3$ / $I_2$, in the second. Therefore, the four zones identified in cases a) and c) of Figure 1 correspond to:

*Zone 1. q(k) > α ∧ q(k') > α ⇒ Contribution to Δ(k', k) >0 (assortativity)*

*Zone 2. q(k) > α ∧ q(k') < α ⇒ Contribution to Δ(k', k) <0 (disassortativity)*

*Zone 3. q(k) < α ∧ q(k') > α ⇒ Contribution to Δ(k', k) <0 (disassortativity)*

*Zone 4. q(k) < α ∧ q(k') < α ⇒ Contribution to Δ(k', k) >0 (assortativity)*

On the other hand, the b term shows a slightly different behavior with the lower left and upper right zones (zones 1 and 4) disconnected. The b term, case b) of Figure 1, is third order in its variables and is not separable in *q(k')* and *q(k)*. Therefore, no analytical solution can be provided for their zeros. However, it is simple to determine that *q(k')*=0 implies *q(k)*= $I_3$ / $I_2$ and vice versa. In other words, zone 4 for the b term intersects the axes at the same values as term c. Now, zones 2 and 3 can merge in a single one. This result, and the previous for terms (a-1) and b, shows that the $I_2$ and $I_3$ coefficients determine the limit between assortativity and disassortativity in equation (19). In particular, low $I_3$ values increase the size of zone 1. Therefore, when a single term is used for $\Delta(k', k)$, $I_2$, and/or $I_3$ define the boundaries between zones without influence of the a, b, or c parameters.

When more than one parameter appears in $\Delta(k', k)$, we can find the intersection of the zero isocontour line with the $q(k')$ and $q(k)$ axes setting equation (19) to zero and making $q(k')$ or $q(k)$ =0. In this form, we observe that when $\Delta(k', k)$ includes only the (a-1) and c terms, or the b and c terms, the c term has no influence. Therefore, the intersection point does not depend on any of the model parameters. However, when the (a-1) and b terms are present, with or without the c term, we obtain for the intersection point,

$$q_0(k) = \frac{-bI_3 \pm \sqrt{I_3[b^2 I_3 + 4(a-1)(a-1+bI_2)]}}{2(1-a-bI_2)} \qquad (20)$$

Equation (20) shows that, in the general case, the boundaries between assortative and disassortative zones depend on a, b, $I_2$, and $I_3$ but not on c.

The isocontour lines in the three cases of Figure 1 show that the gradient is higher in zone 1, with zones 2, 3, and specially 4, being much more flat. Zone 4 exhibits a shallow maximum whereas in zones 1, 2 and 3 $\Delta(k', k)$ increases (in absolute value) uniformly. The maximum in zone 4 is found, see equation (19), at $q(k)=q(k')= \sqrt{(I_3/3)}$, for the (a-1) term, and at $q(k)=q(k')= (2/3) (I_3/I_2)$ for the c term.

When comparing the (a-1) and b terms in Figure 1, the first one, which has the $q(k) \cdot q(k')$ leading term, exhibits slightly higher values in zone 1 and values about ten times greater in zones 2, 3 and 4. In turn, term c has values about ten times smaller than term b in the four zones. Therefore, in general, the (a-1) term can be considered as the reference term, with the b and the c ones representing a correction. The b and c values define the strength and sign of the correction. In other words, the a, b, and c parameters determine how strongly the inner correlation structure reflects in the network. The degree correlation trend shown in Figure 1 inverts for a<1 or b, c<0.

**III. EXPERIMENTAL VALIDATION**

We can use information from real networks to validate the suitability of the previous treatment for the analysis of the assortative or disassortative behavior. Thus, we have considered a test set of 15 different modeled, social, natural, and technological networks. The different considered cases are collected in Table II.

For the different networks, we compute *P(k)* and *P(k', k)* directly from the network topology as [25],

$$P(k) = \frac{N_k}{N}$$
$$P(k',k) = \frac{N_{k',k}}{2E}$$
(21)

where N is the number of nodes, E the number of edges, $N_k$ the number of nodes with degree k and $N_{k'\,k}$ the number of edges linking nodes of degree k' with nodes of degree k. *q(k)* is computed from *P(k)* using equation (10). For assessing the global degree correlation, Newman's $r_c$ global degree correlation coefficient is computed with equation (1). All the probability distribution calculations have been carried out with the ProNet software package [45]. The $r_c$ coefficient is included in Table II which shows that $r_c$ varies in a range from -0.1984 (Internet) to +0.2679 (High Energy Physics Theory collaboration).

Using *P(k', k)* we can compute *Δ(k', k)* and check the behavior predicted by equation (19). Thus, we have collected in Figure 2 four different cases corresponding to representative kinds of networks, selected from the ones in Table II. The first one, case a), is the modeled Barabási-Albert network. This is generated with the 0.8.1-beta version of the Gephi networks manipulation and visualization package [46]. The second example, case b), is a technological network, the internet routers structure [35]. In turn, case c) depicts an example of a biological network, the diseasome network [43]. Finally, case d) shows a social network, Slashdot [36].

The first thing to note is that the four cases in Figure 2 show the basic four zones structure for $\Delta(k', k)$ predicted by equation (19) and depicted in Figure 1. Thus, in case a), we can identify zones 2, 3 and 4. This is consistent with a behavior given essentially by the b term of equation (19), see case b) of Figure 1. The sign of the isocontour lines is reversed with respect to the three cases of Figure 1. This should correspond to a negative value of the dominant coefficient in equation (19). In this case, zone 1 is not appreciable because the zone 1/zone 4 gap places it out of the $q(k')$ and $q(k)$ value range. Cases b and c exhibit four sections that can be directly mapped to the four zones predicted by equation (19). As in case a) the sign of the isocontour lines is reversed with respect to the values in Figure 1. Case b) shows slightly separated zones 1 and 4, whereas in case c) both zones merge. In both cases, zones 2 and 3 exhibit small positive values. Finally, case d) shows again the four zones with a small zone 4 and isocontour line signs following the pattern in Figure 1.

The ability of the models represented by equation (18) to quantitatively describe the assortative or disassortative behavior of the considered networks, is analyzed by applying a multilinear regression to the $P(k', k)$, $q(k)$, and $q(k')$ data. In this form, the most significant a, b, and c values are determined. As the summation used to compute the ANND distribution [23], this fitting procedure has the effect of smoothing the possible fluctuations in the experimental $P(k', k)$ values. In all cases, as a basic index for the goodness of the fit, we use the adjusted coefficient of determination, $R^2_{adj}$. This represents the proportion of the variance of the independent variable predicted by the dependent ones, taking into account the number of parameters with respect to the number of observations. To select the most significant combination of fitted a, b, and c parameters we apply a stepwise regression. Here, we select a null hypothesis probability limit of 0.05. Thus, we include in the regression function the parameter with the smaller F probability below that threshold. In turn, parameters already in the regression are removed if their F probability becomes larger than 0.10 due to the inclusion of the new parameter. From the series of models generated in the process we select the simplest one where the upper limit for the coefficient of determination variation is 0.01 and the standard deviation variation is at most 10%. The $I_2$ and $I_3$ values for each network and the fitted a, b, and c parameters as well as the $R^2_{adj}$, are collected in Table II. All the calculations are performed with version 21 of the SPSS statistical package.

In all cases, we find that the null hypothesis of each regression coefficient being zero, computed through the F probability, is smaller than $10^{-3}$. This is well below the significance level limit of 0.05. In addition, the standardized adjusted regression coefficients show that the a term [see equation (18)] has the higher weight in the regressions, except for the Barabási-Albert (BA) network, where the b term is the most important, and the last case of Table II where only the c term is present. The fact that the b term has the most weight in the BA network agrees with the shape of its $\Delta(k', k)$ distribution [see case a) of Figure 2] where zone 4 follows the pattern of the b term shown in case b) of Figure 1.

The $R^2_{adj}$ results show that the fitted models are able to describe at least 90% of the observed variation of $P(k', k)$ except in three cases: Wikivote ($R^2_{adj}$=0.856), High Energy Physics Theory collaboration ($R^2_{adj}$=0.875) and *S. cerevisiae* protein interaction network ($R^2_{adj}$=0.782). In addition, it is interesting to note that the simplest model, consisting of just the a term, yields very good regressions with $R^2_{adj}$ values smaller than those in Table II just by 0.017 in the worst case (Pretty Good Privacy network). So, even this simple model is able to provide an overall description of the degree-correlation behavior.

The variation of $P(k', k)$ as a function of $q(k')$ and $q(k)$ is shown in Figure 3 for the experimental and regression generated data in some selected cases. Thus, in case a) we have the designed (BA) network included in Table II. As shown in Table II the regression expression exhibits a very good fit with $R^2_{adj}$ =0.998. We can observe that the regression generated data are in good agreement with the experimental ones. The general variation is well reproduced, including the change from convexity to concavity as the value of the isocontour lines increases. Case b) of Figure 3 shows an actual network: the telephony contacts network collected in Table II. Here, the regression, $R^2_{adj}$=0.900, corresponds again to a good fit of the experimental data to the mathematical model. The figure shows the experimental data to exhibit a certain irregular, noisy, pattern. The regression isocontours follow the same trend that the experimental data but offer a smoother perspective. Finally, case c) of Figure 3 corresponds to the worst regression case: the *S. cerevisiae* protein interaction network which has a $R^2_{adj}$=0.782. We observe now that the experimental data are much more irregular, with a central zone resembling a random pattern. This lack of regularity can be the source of the poor regression results. In fact, case c) of Figure 3 shows how the fitted expression is only able to roughly represent the general *P(k', k)* variation.

It is possible to compare the assortative or disassortative character given by the $r_c$ coefficient with the values of the a, b, and c parameters. So, we consider all cases where the $r_c$ value is significant (greater than 0.1 in absolute value). We observe that when the a term is dominant in the model, a>1 cases correspond to assortative networks and a<1 to disassortative ones. Equation (19) and Figure 1 shows that this is due to the sign of the dominant part in the a term (zone 1). In the case of the *S. cerevisiae* protein interaction network only the c term appears. Therefore, the zero isocontour lines intersect the *q(k')* and *q(k)* axes at $I_3/I_2$=0.0083 within a full axes range of 0.0 to 0.015. The disassortative character of the network is due, then, to the large contribution of zones 2 and 3, see Figure 1 case c).

In addition, the dependence of the degree-degree correlation coefficient $r_c$ and the $I_2$ and $I_3$ indices is also considered and shown in Figure 4. While no relationship is found between $r_c$ and $I_2$ or $I_3$, a clear smooth variation of $r_c$ with both $I_2$ and $I_3$ is observed. Therefore, based on the previous results, only $I_3$ is necessary to provide a reasonable description of degree correlation in a single network using the a term of equation (18). However, as shown here, models including $l_2$ and $l_3$ need to be considered to model degree correlation variations among different networks.

## IV. CONCLUSIONS

This paper presents a general treatment of degree correlation variations in complex networks. A function, $\Delta(k, k')$, measuring the deviation from the uncorrelated case is used to map out individual degree-degree correlations within networks. A graphical representation of $\Delta(k, k')$ permits visualizing the (dis)assortativity variation inside a network.

Use of the total degree probability distribution allows application of group theory to a Taylor expansion of the quotient between the joint probability function, $P(k', k)$, and the independent events probability function. This treatment permits deriving expressions for $P(k', k)$ of arbitrary complexity. Limiting the series expansion to third order we obtain an expression depending on three parameters that represents a family of seven different models.

Use of this expression to compute $\Delta(k, k')$ permits us to predict a nonhomogenous correlation degree structure in networks. Four different zones are identified, defining two assortative and two disassortative regions in a network. In the individual terms of the expression, the boundaries between assortative and disassortative regions are given by the $l_2$ and $l_3$ network topological indices. In the general case, the parameters of the expression also affect the extension and location of the four correlation regions. The magnitude and sign of these parameters determine how the inner correlation structure reflects in the network.

The predicted behavior is tested with experimental data from 15 different modeled, technological, social, and biological networks. Using one representative case of each category, the experimental $\Delta(k, k')$ function is determined. The results show that the four degree correlation zones predicted by the model are actually present in the considered cases. Therefore, this specific, nonhomogeneous correlation structure seems to be a characteristic of real networks.

The ability of the proposed model to quantitatively describe *P(k', k)* is determined by applying a multilinear regression to the experimental data. The regression procedure determines the most significant combination and the value of the parameters in the model. The results show that in 12 of the 15 cases the model describes more than 90% of the *P(k', k)* variance. However, in even the worst case, *S. cerevisiae* protein interaction network, the *P(k', k)* variance is described in more than 78%. An interesting observation is that the simplest model possible, the one with the parameter corresponding to the lower order term, only decreases the ability to describe the P(k', k) variation in 1.7 % in the worst case. These results show that the present approach can be a useful way of representing degree correlation variations in a network.

Comparison of Newman's degree correlation coefficient with the lower order term of the *P(k', k)* expression shows that its coefficient plays a leading role in determining the (dis)assortative character of the network.

When considering the set of actual networks, no direct relationship is found between the Newman's degree-degree correlation coefficient, $r_c$, and the $I_2$ or the $I_3$ topological indices. However, there exists a smooth variation of $r_c$ with $I_2$ and $I_3$. This shows that, to describe correlation variations over different networks, a correlation model needs to incorporate both indices.

The model here presented, can be useful for two purposes. First, given a network it provides a compact mathematical expression describing the inner variation of degree-degree correlation. Second, given a degree correlation function it permits to define the desired correlation behavior just by tuning the parameter(s) in the expression. This allows the building of networks with a specific, predefined, behavior.

**ACKNOWLEDGEMENTS**

A. N. wishes to thank Prof. A. L. Barabási for his helpful suggestions while visiting the Center for Complex Networks Research (CCNR) as well as the assistance and support of all the members of the CCNR. The authors also wish to thank Dr. P. Hövel for providing the anonymized mobile-phone data and the Universidad de Castilla-La Mancha for financial support.

**Table I.** Character table for the $G_2$ interchange group. $a_1$ and $a_2$ identify the totally symmetric and antisymmetric irreducible representations, respectively.

| $G_2$ | $\hat{E}$ | $\hat{I}$ |
|---|---|---|
| $a_1$ | 1 | 1 |
| $a_2$ | 1 | -1 |

**Table II.** The set of 15 networks considered in this study. The table includes the number of nodes, N, number of edges, E, and Newman's correlation coefficient for end nodes of edges defined in equation (1), $r_c$. The $I_2$ and $I_3$ indices, the adjusted coefficient of determination, $R^2_{adj}$, and the a, b and c regression parameters, see equation (18), are also included.

| Network | N | E | $r_c$ | $I_2$ | $I_3$ | $R^2_{adj}$ | a | b | c |
|---|---|---|---|---|---|---|---|---|---|
| BA Network | 50000 | 49999 | -0.0315 | 0.1596 | 0.0430 | 0.998 | 0.318 | 4.743 | -87.615 |
| US Power Grid[a] | 4941 | 13188 | 0.0035 | 0.1613 | 0.0337 | 0.942 | 1.079 | ---- | ---- |
| US Airports[b] | 1574 | 28236 | -0.1216 | 0.0065 | $5.37 \cdot 10^{-5}$ | 0.935 | 0.981 | ---- | ---- |
| Internet[c] | 22963 | 48436 | -0.1984 | 0.0580 | 0.0090 | 0.998 | 0.675 | 8.000 | ---- |
| Slashdot[d] | 82168 | 948464 | -0.0476 | 0.0063 | 0.0001 | 0.999 | 1.046 | ---- | ---- |
| Enron[e] | 36692 | 367662 | -0.1108 | 0.0107 | 0.0003 | 0.985 | 0.978 | 4874.606 | ---- |
| Gnutella[f] | 62586 | 147892 | -0.0926 | 0.0702 | 0.0074 | 0.950 | 0.910 | ---- | ---- |
| WikiVote[g] | 7115 | 1036801 | -0.0686 | 0.0042 | $2.10 \cdot 10^{-5}$ | 0.856 | 1.010 | ---- | ---- |
| Pretty Good Privacy[h] | 10680 | 24340 | 0.2394 | 0.0368 | 0.0022 | 0.980 | 2.746 | -79.377 | 3621.932 |
| Telephony contacts[i] | 3604259 | 6183409 | 0.2549 | 0.1025 | 0.0128 | 0.900 | 1.671 | -5.976 | ---- |
| Amazon[j] | 262111 | 1234877 | -0.0025 | 0.0757 | 0.0074 | 0.925 | 0.873 | 259.321 | ---- |
| High Energy Physics Theory collaboration[k] | 9877 | 51971 | 0.2679 | 0.0419 | 0.0023 | 0.875 | 1.235 | ---- | ---- |
| Diseasome[l] | 2821 | 2673 | -0.1387 | 0.1798 | 0.0490 | 0.997 | 0.612 | 2.098 | ---- |
| *H. sapiens* protein interaction[m] | 13242 | 74288 | -0.0826 | 0.0117 | 0.0002 | 0.942 | 0.822 | ---- | 9079.060 |
| *S. cerevisiae* protein interaction[m] | 6052 | 90252 | -0.1182 | 0.0066 | $5.48 \cdot 10^{-5}$ | 0.782 | ---- | ---- | 0.547 |

[a] US Power Grid [34]. Downloadable from Tore Opsahl web page: http://toreopsahl.com/datasets/

[b] Complete US airport network in 2010. Downloadable from Tore Opsahl web page: http://toreopsahl.com/datasets/

[c] Internet routers structure at July 22, 2006 [35]

[d] Slashdot Zoo social network [36]. Downloadable from the Stanford large network dataset collection: http://snap.stanford.edu/data/#socnets

[e] Enron email network [36, 37]. Downloadable from the Stanford large network dataset collection: http://snap.stanford.edu/data/#socnets

[f] Snapshot of the Gnutella peer-to-peer file sharing network from August 2002 [38, 39]. Downloadable from the Stanford large network dataset collection: http://snap.stanford.edu/data/#socnets

[g] Wikipedia adminship voting data from the inception of Wikipedia until January 2008 [40]. Downloadable from the Stanford large network dataset collection: http://snap.stanford.edu/data/#socnets

[h] List of edges of the giant component of the network of users of the Pretty Good Privacy algorithm for secure information interchange [41]. Downloadable from: http://deim.urv.cat/~aarenas/data/welcome.htm

[i] Reciprocal contacts (calls+text messages) between telephony users along February, 2009. Anonymized call data from a single mobile-phone provider in an industrialized country. Data courtesy of P. Hövel.

[j] Amazon product copurchasing network from March 2, 2003 [42]. Downloadable from the Stanford large network dataset collection: http://snap.stanford.edu/data/#socnets

[k] Collaboration network of arXiv High Energy Physics Theory [38]. Downloadable from: the Stanford large network dataset collection: http://snap.stanford.edu/data/#socnets

[l] Human diseasome bipartite graph [43]. Downloadable from: http://www.barabasilab.com/pubs/CCNR-ALB_Publications/200705-14_PNAS-HumanDisease/Suppl/index.htm

[m] Homo sapiens and *Saccharomyces cerevisiae* protein interaction networks. Data from June 28, 2011. Downloaded from the Protein Interaction Network Analysis (PINA) platform [44]: http://cbg.garvan.unsw.edu.au/pina/

**Figure 1.** Isocontour maps of the Δ*(k', k)* correlation function components, given in equation (19), as a function of the *q(k')* and *q(k)* probability distributions. The axes include the variation of the *k'* and *k* degrees. Case a) (a-1) term. Case b) b term. Case c) c term. Interval between isocontour lines 0.02 in cases a) and b) and 0.002 in case c). For reference purposes, case a) includes a 0.001 isocontour line in zones 1 and 4, case b) a -0.001 isocontour line in zones 2 and 3, and case c) a -0.0001 isocontour line in zones 2 and 3. Darker zones correspond to lower Δ(*k', k*) values.

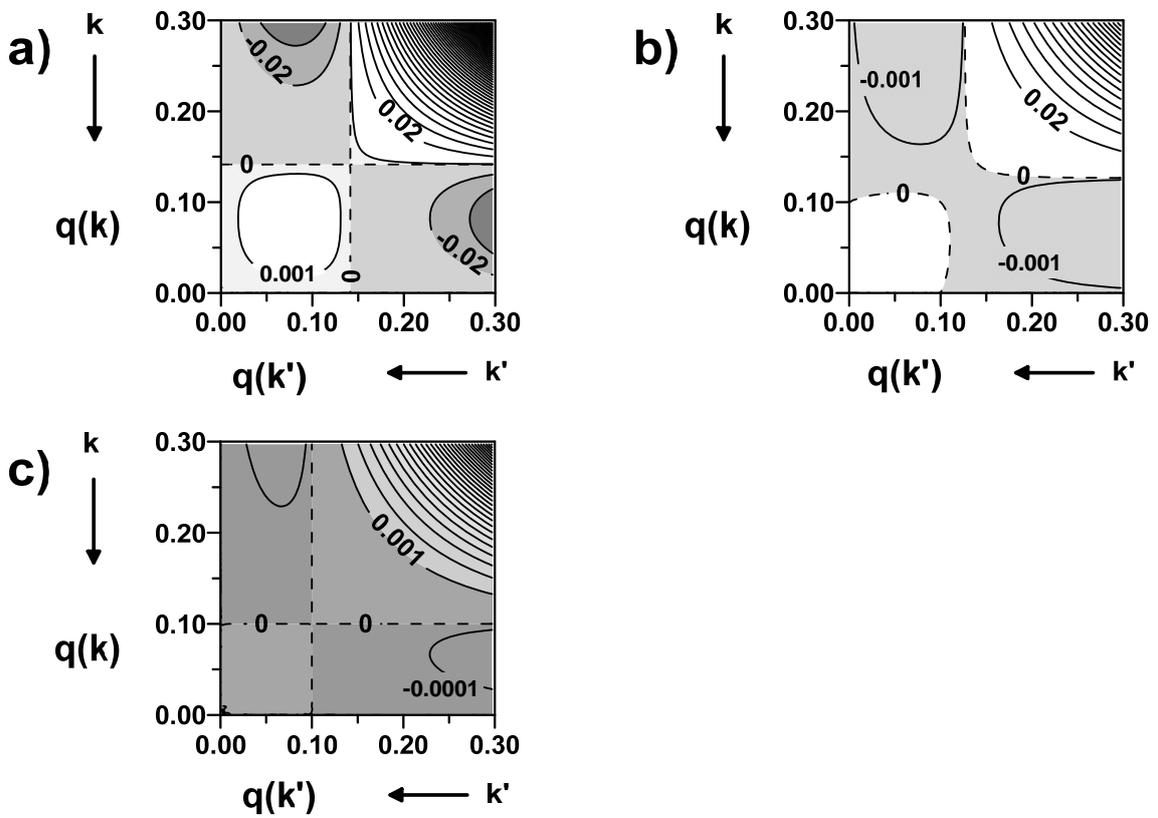

**Figure 2.** Δ(k', k) correlation function for representative cases of modeled, technological, social and biological networks. Case a) Barabási-Albert generated network. Case b), internet routers structure network. Case c), Slashdot social network. Case d), human diseasome network. Interval between isocontour lines 0.002. Darker zones correspond to lower Δ(k', k) values.

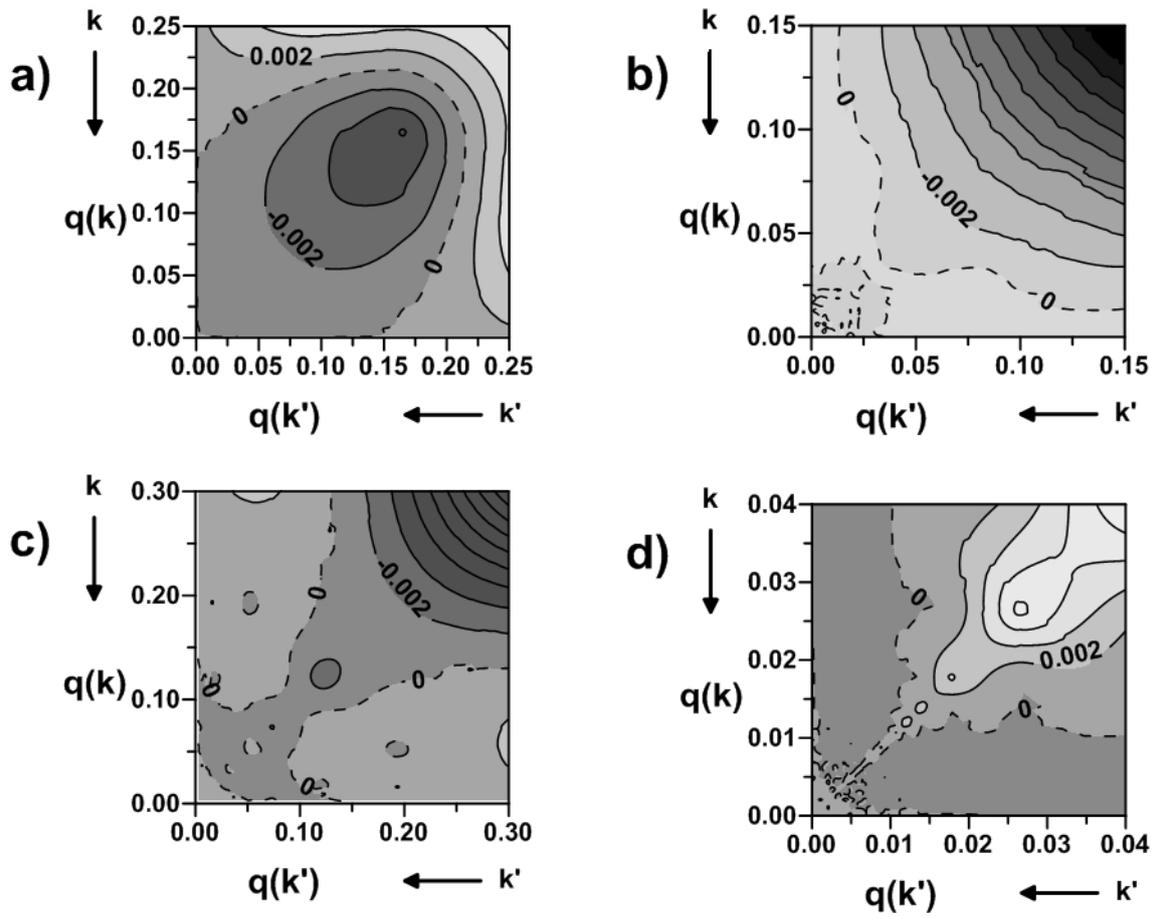

**Figure 3.** Graphical comparison of experimental (left diagrams) versus computed (right diagrams) $P(k', k)$ variation. Case a), Barabási-Albert network. Interval between isocontour lines 0.005. Case b), telephony contacts network. Interval between isocontour lines 0.002. Case c), *S. cerevisiae* protein interaction network. Interval between isocontour lines $5.0 \cdot 10^{-5}$. Darker zones correspond to lower $\Delta(k', k)$ values. The three cases considered are included in Table II.

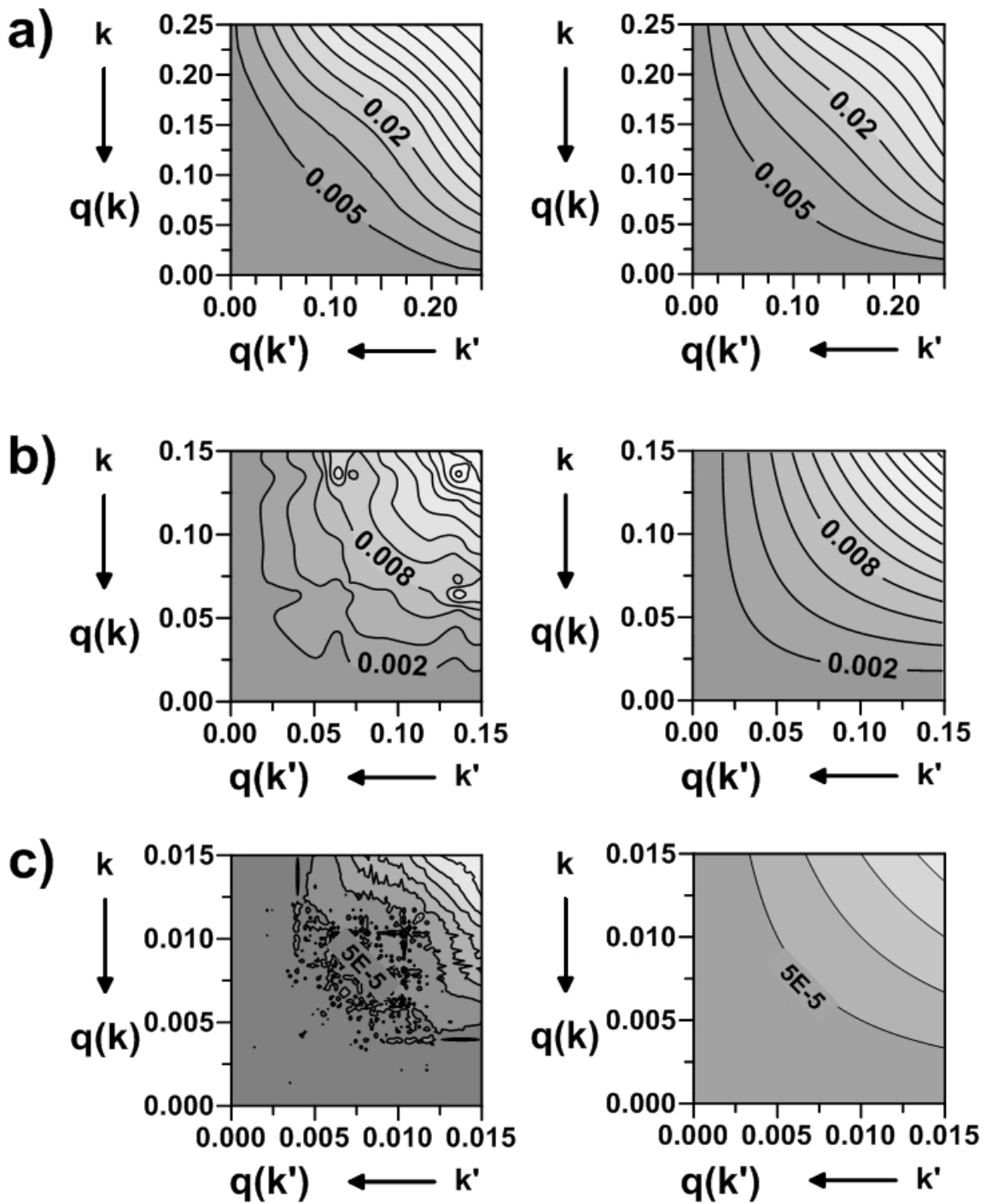

**Figure 4.** Isocontour map of Newman's $r_c$ global degree correlation coefficient as a function of the $I_2$ and $I_3$ network indices. Interval between isocontour lines 0.05. Darker zones correspond to lower $r_c$ values.

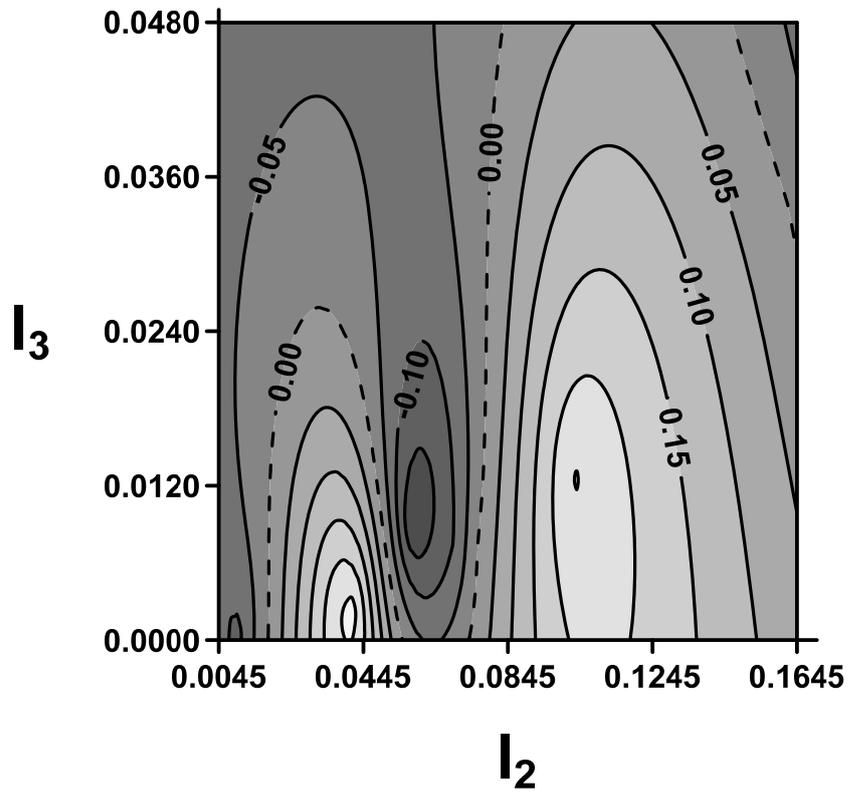